\documentclass{PoS}
\usepackage{amssymb}   % for math
\usepackage{amsmath}   % for math
\usepackage{wrapfig}

\title{A new Bayesian approach to the reconstruction of spectral functions}

\ShortTitle{A new Bayesian approach to the reconstruction of spectral functions}

\author{Yannis Burnier \\
        Institut de Th\'eorie des Ph\'enom\`enes Physiques, Ecole Polytechnique F\'ed\'erale de Lausanne, CH-1015, Lausanne, Switzerland,\\
        Albert Einstein Center, University of Bern, Sidlerstr. 5, CH-3012 Bern, Switzerland\\
        E-mail: \email{yannis.burnier@epfl.ch}}
        
\author{\speaker{Alexander Rothkopf} \\
        Albert Einstein Center, University of Bern, Sidlerstr. 5, CH-3012 Bern, Switzerland\\
        E-mail: \email{rothkop@itp.unibe.ch}}

\abstract{We present a novel approach for the reconstruction of spectra from Euclidean correlator data that makes close contact to modern Bayesian concepts. It is based upon an axiomatically justified dimensionless prior distribution, which in the case of constant prior function $m(\omega)$ only imprints smoothness on the reconstructed spectrum. In addition we are able to analytically integrate out the only relevant overall hyper-parameter $\alpha$ in the prior, removing the necessity for Gaussian approximations found e.g. in the Maximum Entropy Method. Using a quasi-Newton minimizer and high-precision arithmetic, we are then able to find the unique global extremum of $P[\rho|D]$ in the full $N_\omega\gg N_\tau$ dimensional search space. 
The method actually yields gradually improving reconstruction results if the quality of the supplied input data increases, without introducing artificial peak structures, often encountered in the MEM. To support these statements we present mock data analyses for the case of zero width delta peaks and more realistic scenarios, based on the perturbative Euclidean Wilson Loop as well as the Wilson Line correlator in Coulomb gauge.}

\FullConference{31st International Symposium on Lattice Field Theory LATTICE 2013\\
		 July 29th - August 3rd, 2013\\
		 Mainz, Germany}

\begin{document}

\section{Motivation}

Lattice QCD constitutes an important tool in elucidating the physics of the strong force from first principles. Its Monte-Carlo based simulation techniques in Euclidean time allow us to investigate a multitude of non-perturbative settings out of reach from any other method available today. 
To connect to phenomenology, real-time information however needs to be extracted, whose direct determination remains out of reach due to the presence of the notorious sign problem.

All information encoded in the two-point functions of a quantum field theory in Minkowski time can be obtained from the real-valued spectral function $\rho(\omega)$ of the system. I.e. it allows us to generate any propagator, corresponding to one of the possible operator orderings, by an integration in the complex plane. For the Euclidean correlator on the other hand the connection to $\rho(\omega)$ is established via a real-valued integral kernel
\vspace{-0.2cm}\begin{align}
 D(\tau)=\int d\omega K(\tau,\omega) \rho(\omega).\label{Eq:ConvCont}
\end{align}

\vspace{-0.2cm}Lattice QCD simulations give us access to stochastic estimates of $D(\tau_i)=D_i$ at a finite number $N_\tau$ of points along the temporal axis. The task at hand is to give meaning to the inherently \textit{ill-defined} problem of extracting from this limited and noisy data the rich features of $\rho(\omega_l)=\rho_l>0$ along $N_\omega\gg N_\tau$ frequencies.

The most common approach to invert a relation such as \eqref{Eq:ConvCont} is based on Bayesian inference \cite{Jarrell:1996} (for a recent non-Bayesian proposal see e.g.~\cite{Cuniberti:2001hm}). Beginning with Ref.~\cite{Asakawa:2000tr} the strategy of determining the most probable spectral functions based on Bayes theorem 
\begin{align*}
 P[\rho|D,I]=\frac{P[D|\rho,I]P[\rho|I]}{P[D|I]} \quad\Rightarrow\quad \left.\frac{\delta P[\rho|D,I]}{\delta \rho}\right|_{\rho=\rho^{Bayes}}=0 \qquad \begin{array}{cl}P[D|\rho,I] & {\rm likelihood\,probability}\\ P[\rho|I] & {\rm prior\,probability} \\ P[D|I] & {\rm evidence} \end{array}
\end{align*}
has found its way into the lattice QCD community under the name of Maximum Entropy Method (MEM), where the prior probability $P_{MEM}[\rho|I]={\rm exp}[S_{SJ}]$ is given by the Shannon-Jaynes entropy \cite{Jarrell:1996,Asakawa:2000tr}. While important insight on finite temperature QCD \cite{Asakawa:2003re} has been gained using the MEM, several technical short-comings (which have afflicted our own studies in the past \cite{Rothkopf:2011db}) prevented us from utilizing the full potential of the Bayesian approach:
\begin{itemize}
 \item The dimensionality of the search space in the standard implementation by Bryan is fully determined by the number of datapoints $N_\tau$. If the structures encoded in the spectrum are spread over a wide frequency range it has been shown \cite{Rothkopf:2011} that the limited number of basis functions can adversely impact the reconstruction result. One consequence is that comparisons of spectra obtained at different temperatures (i.e. from datasets with different $N_\tau$) contain additional systematic uncertainty, as the underlying search space differs.
 \item If the search space is decoupled from the number of datapoints \cite{Rothkopf:2011} and the number of available basis functions increased, we find that the convergence of the underlying optimization task slows down significantly, since there exist flat directions in the Shannon-Jaynes entropy functional ($\rho_l,m_l \ll 1/\alpha$ or $\rho_l\ll m_l$). In addition, a multitude of peaks appears that are not part of the actual encoded spectrum and it is not trivial to identify them as such.
 \item The MEM contains a hyperparameter $\alpha$, which is treated self-consistently based on a Gaussian approximation \cite{Jarrell:1996}. This can be justified only if the default model $m(\omega)$ lies close to the correct spectrum. In practice, we usually do not possess reliable information about the peak structures we wish to reconstruct, so that a different treatment of $\alpha$ is called for.
\end{itemize}

Here we wish to report on a novel strategy \cite{Burnier:2013nla} for spectral function reconstruction that attempts to remedy the above mentioned issues. In Section:
\begin{enumerate}
 \setcounter{enumi}{1}
 \item We introduce a modified likelihood function that contains additional knowledge on how neutral reconstructions can be achieved.
 \item A prior distribution for $\rho$ is constructed axiomatically, which in the presence of a constant default model $m(\omega)=m_0$ imprints only smoothness on the reconstructed spectrum.
 \item Through the advantageous analytical structure of this new prior functional we are able to integrate out explicitly the hyperparameter $\alpha$ from the joint probability distribution.
 \item We present spectral reconstructions from mock data containing a three delta peak spectrum as well as a challenging scenario based on the hard-thermal loop Wilson loop and Wilson lines in Coulomb gauge \cite{Burnier:2013fca,Burnier:2013lat}. Using the quasi-Newton LBFGS algorithm allows us to carry out the necessary optimization tasks in the full $N_\omega$ dimensional search space. 
\end{enumerate}

\section{The likelihood}

The datapoints $D_i$ we use to reconstruct the spectrum are assumed to arise from the averaging of Gaussian distributed measurements. Consequently the probability of the data, given a test spectral function $\rho$ can be quantified using the quadratic distance
\begin{align}
L=\frac{1}{2}\sum_{ij}(D_i-D^\rho_i)C^{-1}_{ij}(D_j-D^\rho_j),\label{Eq:LFunc}
\end{align}
where $C_{ij}$ denotes the correlation matrix of the data and $D^\rho_i$ are the values obtained by plugging $\rho$ into the discretized version of Eq.~\eqref{Eq:ConvCont}
\vspace{-0.2cm}\begin{align}
D^\rho_i=\sum_{l=1}^{N_\omega}\, \Delta\omega_l\, K_{il} \rho_l. \label{Eq:ConvolutionDiscr}
\end{align}
Note that in the function $L$ the discretization of frequencies $N_\omega$ and $\Delta\omega_l=\omega_{l+1}-\omega_{l}$ does not enter explicitly, as it operates on the level of Euclidean time datapoints. Furthermore we know that if $L\gg N_\tau$, $\rho$ does not reproduce the measured data within errorbars. Contrary, if $L\ll N_\tau$, it is to be expected that the inevitable noise in the data is overfitted and artificial distortions in the reconstruction ensue. Actually, if we were to know the correct spectral function underlying the Gaussian distributed measurements $D_i$ and plug it into Eq.~\eqref{Eq:LFunc} via Eq.~\eqref{Eq:ConvolutionDiscr}, we will obtain $L\approx N_\tau$. Hence we require for the most neutral reconstruction that $L=N_\tau$ by adding a constraint to the function $L$
\vspace{-0.5cm}\begin{align}
 P[D|\rho,I]={\rm exp}[-L - \gamma(L-N_\tau)^2],\label{PDrhoI}
\end{align}
where we take $\gamma\to\infty$ numerically. We have not yet achieved to regularize the procedure. Indeed there still exist an infinite number of degenerate solutions maximizing Eq.~\eqref{PDrhoI}. From Bayes theorem we know that taking into account prior knowledge is the key to a unique answer.

\section{A new prior distribution}

The prior distribution used in the MEM has been derived, based on very general axioms that were chosen with image reconstruction in mind \cite{Jarrell:1996}. In our case the focus is quite different. Besides emphasizing the faithful reconstruction of the structures actually encoded in the data, we aim at preventing the appearance of artificial peaks, which otherwise impede the physics interpretation of the obtained results. Hence we wish our prior probability to favor smooth functions if no other prior information is given\footnote{Reconstructing delta peaked spectra that arise at zero temperature is nevertheless possible. The prior only imprints its signature on those parts of the spectrum that are not constrained by the data. Indeed very good data (large $N_\tau$ and small $\Delta D$) can and will override its  influence (see e.g. Sec.~\ref{Sec5}).}.

To construct $P[\rho|I]={\rm exp}[S]$, we begin with the axiom \textit{Subset independence} already used in the derivation of the Shannon Jaynes entropy, which tells us that $S$ has to be written as an integral over frequencies. Our new axiom \textit{Scale invariance} is related to the fact that depending on the dimensionality of the correlation function $D(\tau)$ in Eq.~\eqref{Eq:ConvCont} the scaling of $\rho(\omega)$ also differs. Note that the default model $m(\omega)$ must exhibit the same scaling as the correct spectrum. We require that the choice of units does not influence the result of the reconstruction so that $S$ has to be a function of the ratio $\rho/m$ only
\vspace{-0.5cm}\begin{align}
 S=\tilde{\alpha} \int d\omega\;  s\Big(\rho(\omega)/m(\omega)\Big). \label{S2}
\end{align}
A dimensionfull hyperparameter $[\alpha]=1/[\omega]$ is introduced to make the overall expression dimensionless.

We proceed with our third axiom \textit{Smoothness of the reconstructed spectra}. If the default model is constant $m(\omega)=m_0$, we require $S$ to choose a smooth spectrum independent of $m_0$. This is achieved by penalizing spectra, whose values at adjacent frequencies $\omega_1$ and $\omega_2$ differ from each other. I.e. the penalty awarded by $S$ between the situation where the ratios $r_l=\rho_l/m_l$ at the two frequencies are the same $r_1=r_2=r$ and where they differ as $r_1=r(1+\epsilon)$ and $r_2=r(1-\epsilon)$ has to be independent of $r$ and must be the same regardless of $r_1\gtrless r_2$. The corresponding discrete differential equation $2s(r)-s(r(1+\epsilon))-s(r(1-\epsilon))=\epsilon^2C_2$ hence leads us to 
\begin{align}
 S=\tilde{\alpha}\int d\omega\; \Big(C_0-C_1\frac{\rho}{m}+C_2\ln\left(\frac{\rho}{m}\right)\Big).\label{Eq:SAx3}
\end{align}

The final step is taken using the axiom \textit{Maximum at the prior}, which we inherit from the MEM derivation. It establishes the Bayesian meaning of $m(\omega)$ by requiring that $S$ is maximal for $\rho=m$. This fixes all the integration constants in Eq.~\eqref{Eq:SAx3} and we end up with the final expression
\begin{align}
 S=\alpha\int d\omega\; \Big(1-\frac{\rho}{m}+\ln\left(\frac{\rho}{m}\right)\Big),\label{Eq:Sfinal}
\end{align}
where $\alpha>0$. Note that $S<0$ for all $\rho\neq m(\omega)$ and that its behavior around the quadratic minimum $\rho=m$ is identical to $S_{SJ}$. Hence the proof of the uniqueness of the Bayesian reconstruction can be directly carried over from the MEM. The last remaining step requires us to treat the positive hyperparameter $\alpha$ in a Bayesian fashion.

\section{Integration of hyperparameters}

Since we do not have any additional information on the hyperparameter $\alpha$, we wish to integrate it out explicitly from the joint probability distribution $P[\rho,D,\alpha,m,I]$. For notational clarity we will refrain for the moment from writing out the label $I$ for the prior knowledge entering Eq.~\eqref{PDrhoI}, as it is completely independent from $\alpha$. The joint probability can be expanded as
\begin{align}
P[\rho,D,\alpha,m]=P[D|\rho,\alpha,m]P[\rho |\alpha,m]P[\alpha|m]P[m]=P[\alpha |\rho,D,m]P[\rho |D,m]P[D|m]P[m],
\end{align}
where a closer inspection reveals that not only $m$ and $\alpha$ are independent but also $P[D|\rho,\alpha,m]$ does not depend on the hyperparameter, since $\rho$ itself is one of the given quantities. Thus the only $\alpha$ dependence is found in $P[\rho |\alpha,m]$,$P[\alpha |\rho,D,m]$ and $P[\alpha]$. Introducing the integrals with respect to $\alpha$ and setting $P[\alpha]=1$ one obtains
\begin{align}
  P[D|\rho]\int d\alpha P[\rho |\alpha,m]=\int d\alpha P[\alpha |\rho,D,m] P[\rho |D,m]P[D|m].
\end{align}
By definition $\int d\alpha P[\alpha |\rho,D,m]=1$ and we can rearrange the terms to get
\begin{align}
 P[\rho |D,m,I]=\frac{P[D|\rho,I]}{P[D|m,I]}\int d\alpha P[\rho | \alpha,m],\label{Eq:IntegrAlpha}
\end{align}
where the labels $I$ have been restored. This expression constitutes the basis for our reconstruction. We carry out the single remaining $\alpha$ integral for large values of S through a next-to-leading order resummation of logarithms, while at small S a numerical integration can be performed. 

\section{Numerical implementation and mock data tests}
\label{Sec5}
\begin{figure}
\vspace{-0.7cm}
\centering
\hspace{-1cm} \includegraphics[scale=0.25,angle=-90]{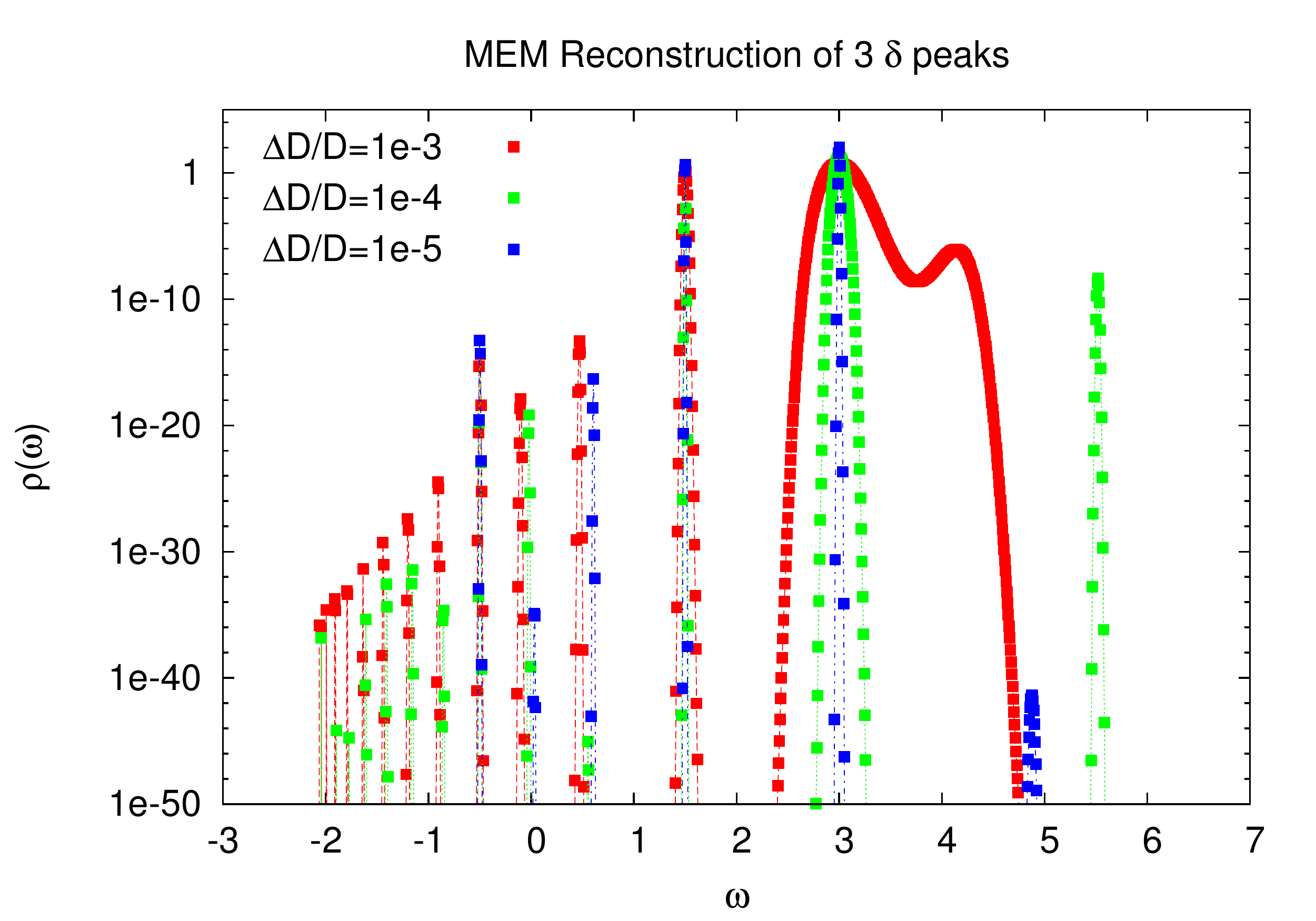}\hspace{0.5cm}
 \includegraphics[scale=0.25, angle=-90]{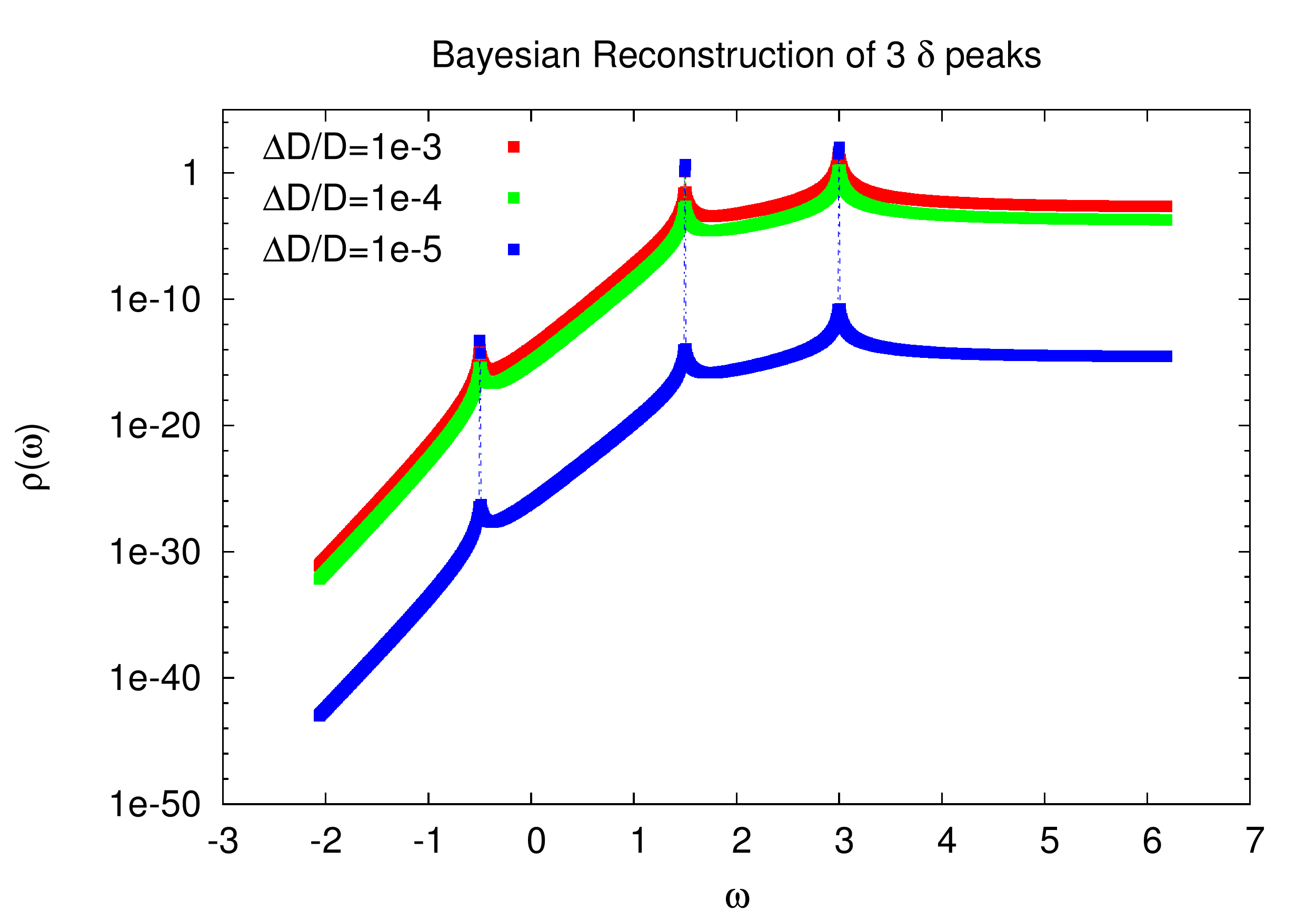}
 \caption{Bayesian reconstruction of a three $\delta$-peak mock spectrum using MEM (left) and our Bayesian method (right). Both methods are supplied $N_\tau=32$ ideal datapoints perturbed by Gaussian noise corresponding to errors of $\Delta D/D=10^{-3}$ (red) $10^{-4}$ (green) $10^{-5}$ (blue). Note the complete absence of artificial peaked structures on the right. At $\Delta D/D=10^{-5}$ there still remains a width of two points (right), since the used $\omega\in[-2,6]$ with $N_\omega=1200$ does not contain the exact values $\omega=-0.5,1.5,3$. }\label{Fig:ThreeDelta}\vspace{-0.3cm}
\end{figure}

To find the maximum of Eq.~\eqref{Eq:IntegrAlpha} we need to vary the $N_\omega\gg N_\tau$ components $\rho_l$ individually. To do so we use the LBFGS algorithm, which approximates the inverse Hessian matrix of the Levenberg-Marquardt scheme iteratively. This significantly reduces the computational cost and we can actually perform reconstructions with $N_\omega\sim{\cal O}(1000)$ without problem. In the following we demonstrate the performance of our method for the kernel $K(\tau,\omega)={\rm exp}[-\omega\tau]$ by mock data analyses. In the first a three delta peak spectrum $\rho(-0.5)=10^{-14},\rho(1.5)=\frac{1}{26},\rho(3)=\frac{25}{26}$ is encoded via Eq.~\eqref{Eq:ConvolutionDiscr} in $N_\tau=32$ datapoints to which Gaussian noise is added. For the reconstruction we discretize $\omega\in[-2,6]$ with $N_\omega=1200$ points and use $512$ bit arithmetic. Assuming no knowledge of the peak positions, we provide a flat prior. As shown in Fig.~\ref{Fig:ThreeDelta}, the results from the MEM (left) and 
our method (right) are quite distinct. For different errors in the data  $\Delta D/D=10^{-3}$ (red) $10^{-4}$ (green) $10^{-5}$ (blue) we consistently obtain smaller reconstructed widths and more importantly manage to completely avoid artificial peak structures that show up in the MEM. 

Where the data constrains the spectrum and overrides the prior, sharp peaks arise. Everywhere else, our prior favors a smooth functional form, which becomes parallel to $m(\omega)$ at large $\omega$. Another important point is that with our new method we find a true maximum of Eq.~\eqref{PDrhoI}. I.e. even for $\Delta D/D=10^{-5}$ the optimizer step size diminishes down to $\Delta=10^{-60}$ within several hours, while in the MEM values of $\Delta<10^{-9}$ are not encountered even after one week of running.

A more demanding test (see also \cite{Burnier:2013nla,Burnier:2013lat}) is based on the analytically calculated hard-thermal loop Wilson loop $W^{HTL}_\square(r,\tau)$ and the Wilson line correlator in Coulomb gauge $W^{HTL}_{||}(r,\tau)$\cite{Burnier:2013fca}. These play a central role in determining the heavy quark potential at finite temperature \cite{Rothkopf:2011db,Burnier:2013nla,Burnier:2013fca,Burnier:2013lat}. In particular the position and width of the lowest lying peak in $\rho(r,\omega)$ at a certain spatial extend $r$ encodes the real- and imaginary part of the $V_{\bar{Q}Q}(r)$ respectively. It is known \cite{Burnier:2013fca} that for the Wilson loop this peak is quite small and embedded in a large background from the cusp divergences of the closed rectangular path. The Wilson lines feature a much smaller background.

\begin{figure}
\vspace{-0.6cm}
\centering
\includegraphics[scale=0.14,angle=-90]{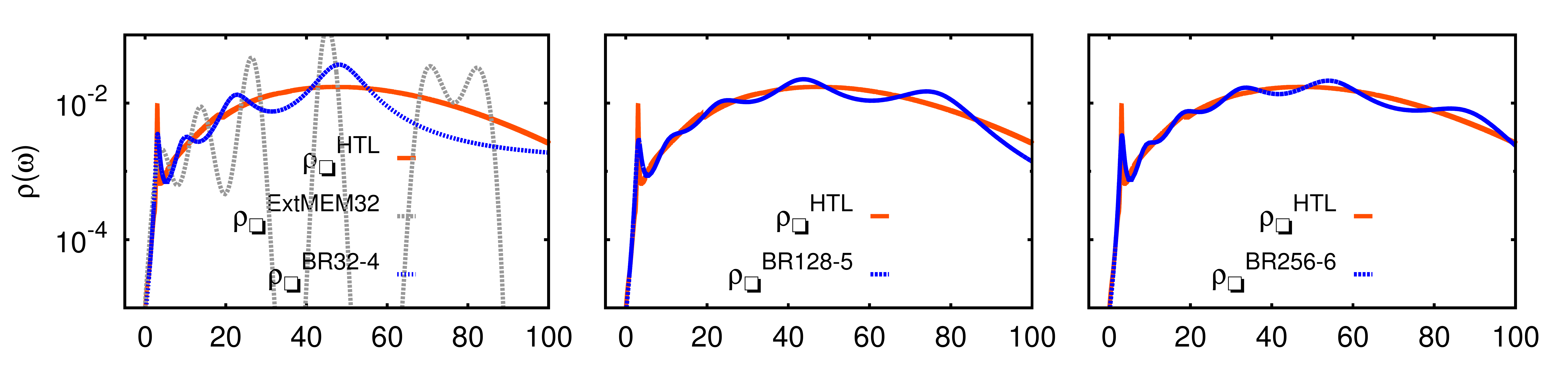}\vspace{-0.2cm}
\includegraphics[scale=0.14, angle=-90]{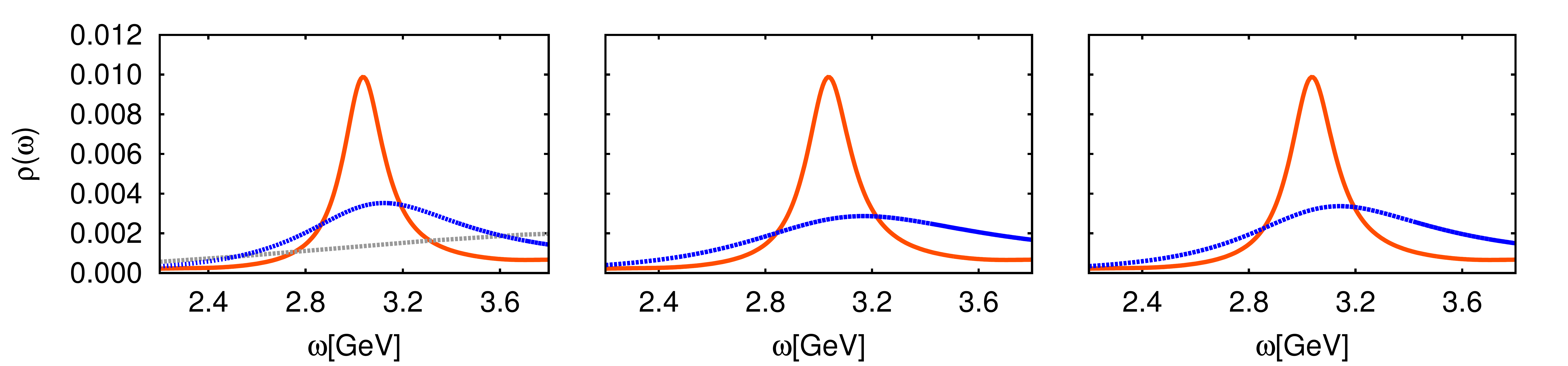}\vspace{-0.1cm}
\caption{Bayesian reconstruction of the $T=2.33\times270{\rm MeV}$ hard-thermal loop Wilson loop spectrum (orange) from $N_\tau=32,128,256$ Euclidean datapoints with $\Delta D/D=10^{-4},10^{-5},10^{-6}$ (left-to-right, blue) at $r=0.264{\rm fm}$. (left column) A reference MEM reconstruction from \cite{Burnier:2013fca} is shown in gray}\vspace{-0.5cm}\label{Fig:Wloops}
\end{figure}

We show in Fig.~\ref{Fig:Wloops} the reconstruction of $\rho^{HTL}_\square(r,\omega)$ (orange) from $W^{HTL}_\square(r,\tau)$ ($N_\omega=1000$, $\omega\in[-126,189]{\rm GeV}$) at $r=0.264{\rm fm}$ ($T=2.33T_C$) for improving input data $N_\tau=32,128,256$ and $\Delta D/D=10^{-4},10^{-5},10^{-6}$ (left-to-right, blue). More points and lower error enable us to better capture the higher frequencies, while the physically important peak remains challenging. Its position at best lies within $10\%$ of the correct value but the width is reproduced only poorly (see also \cite{Burnier:2013lat}).

\begin{figure}
\vspace{-0.7cm}
\centering
\includegraphics[scale=0.14,angle=-90]{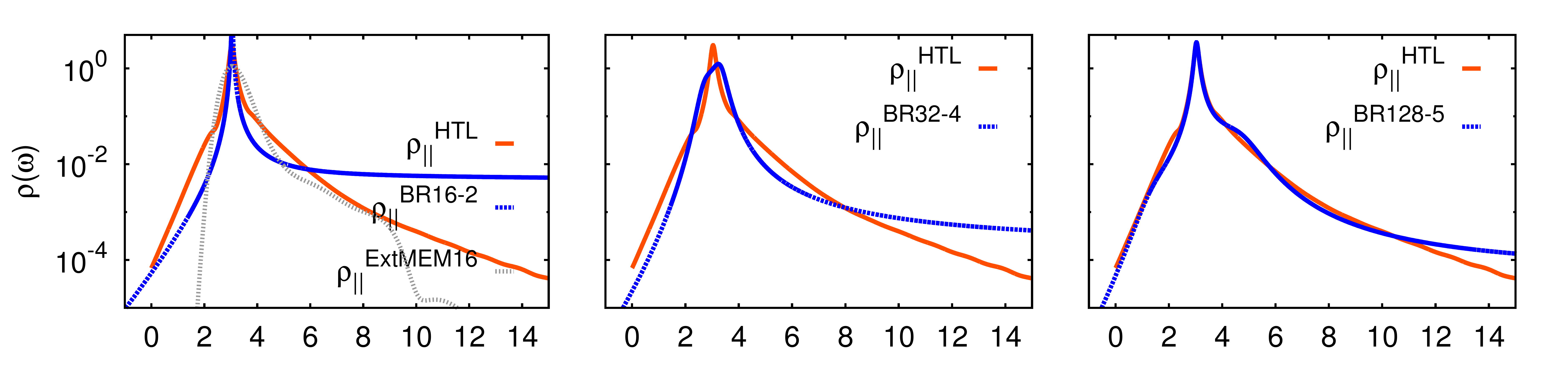}\vspace{-0.25cm}
 \includegraphics[scale=0.14, angle=-90]{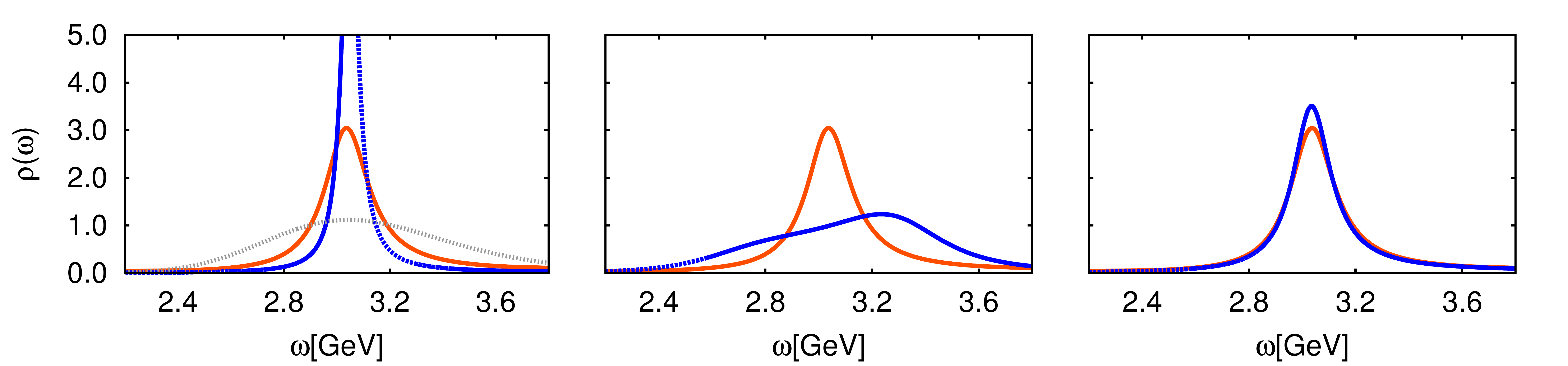}\vspace{-0.1cm}
\caption{Reconstruction of the $T=2.33\times270{\rm MeV}$ hard-thermal loop Wilson line spectrum (orange) from $N_\tau=32,128,256$ Euclidean datapoints with $\Delta D/D=10^{-2},10^{-4},10^{-5}$ (left-to-right, blue) at $r=0.264{\rm fm}$. (left column) A MEM reconstruction with the settings of \cite{Burnier:2013fca}, based on $\Delta D/D=10^{-2}$ data, is shown in gray.}\vspace{-0.4cm}\label{Fig:Wlines}
\end{figure}

In the absence of a divergence induced background, the reconstruction of $\rho^{HTL}_{||}(r,\omega)$ in Fig.~\ref{Fig:Wlines} (orange) performs much better. We show the results from $W^{HTL}_{||}(r,\tau)$ ($N_\omega=1200$, $\omega\in[-15.7,15.7] {\rm GeV}$) for  $N_\tau=16,32,128$ and $\Delta D/D=10^{-2},10^{-4},10^{-5}$ (left-to-right, blue). The position of the peak is already captured satisfactorily with the lowest quality data. Here more of the overall datapoints contribute to the low frequency peak. Thus with $N_\tau=128$ and $\Delta D/D=10^{-5}$, which is still attainable in quenched QCD, we manage to reproduce the width with sub $20\%$ deviation. These results, together with the performance shown in the reconstruction of delta peak structures in Fig.~\ref{Fig:ThreeDelta}, give us confidence that our method can not only contribute to an improved determination of the heavy-quark potential (see \cite{Burnier:2013lat}) but also to the reconstructions of spectral features from lattice simulations in 
general (see e.g. S.~Kim's contribution to this conference \cite{Kim:2013lat}).

We thank T.~Hatsuda, S.~Sasaki, O.~Kaczmarek, S.~Kim, P.~Petreczky and H.T.~Ding for fruitful discussions, C.A.~Rothkopf for insight on Bayesian inference and the DFG-Heisenberg group of Y.~Schr\"oder for generous computer access. This work was partly supported by the Swiss National Science Foundation (SNF) under grant 200021-140234 and the Ambizione grant PZ00P2-142524.

\end{document}